\documentclass[sigconf]{acmart}

\def\BibTeX{{\rm B\kern-.05em{\sc i\kern-.025em b}\kern-.08emT\kern-.1667em\lower.7ex\hbox{E}\kern-.125emX}}

\copyrightyear{}
\acmYear{}
\setcopyright{acmlicensed}
\acmConference[]{}{}{}
\acmBooktitle{}
\acmPrice{}
\acmDOI{}
\acmISBN{}

\usepackage[ruled,vlined]{algorithm2e}
\usepackage{array}
\usepackage{amsfonts}
\usepackage{amsmath}
\usepackage{amssymb}
\usepackage{booktabs}
\usepackage{caption}
\usepackage{comment}
\usepackage{enumitem}
\usepackage{graphicx,epstopdf}
\usepackage{hyperref}
\usepackage{multirow}
\usepackage{natbib}
\usepackage{newtxmath}
\usepackage{subcaption}

\graphicspath{ {figures/} }

\newcommand{\eg}{\emph{e.g.},\xspace}

\newcommand{\ie}{\emph{i.e.,}\xspace}

\newcommand{\etal}{\emph{et~al.}\xspace}
\newcolumntype{x}[1]{>{\raggedright\arraybackslash}p{#1}}

\SetCommentSty{mycommfont}
\renewrobustcmd{\bfseries}{\fontseries{b}\selectfont}
\newrobustcmd{\B}{\bfseries}

\makeatletter 
\g@addto@macro{\@algocf@init}{\SetKwInOut{Output}{Output}} 
\makeatother

\begin{document}

\title{Graph Neural Networks with Continual Learning for Fake News Detection from Social Media}

\author{Yi Han \quad Shanika Karunasekeran \quad Christopher Leckie}
\email{{yi.han, karus, caleckie}@unimelb.edu.au}
\affiliation{%
  \institution{School of Computing and Information Systems, The University of Melbourne}
  \streetaddress{}
  \city{}
  \state{}
  \postcode{}
}

\begin{abstract}
Although significant effort has been applied to fact-checking, the prevalence of fake news over social media, which has profound impact on justice, public trust and our society as a whole, remains a serious problem. In this work, we focus on propagation-based fake news detection, as recent studies have demonstrated that fake news and real news spread differently online. Specifically, considering the capability of graph neural networks (GNNs) in dealing with non-Euclidean data, we use GNNs to differentiate between the propagation patterns of fake and real news on social media. In particular, we concentrate on two questions: (1) Without relying on any text information, e.g., tweet content, replies and user descriptions, how accurately can GNNs identify fake news? Machine learning models are known to be vulnerable to adversarial attacks, and avoiding the dependence on text-based features can make the model less susceptible to the manipulation of advanced fake news fabricators. (2) How to deal with new, unseen data? In other words, how does a GNN trained on a given dataset perform on a new and potentially vastly different dataset? If it achieves unsatisfactory performance, how do we solve the problem without re-training the model on the entire data from scratch, which would become prohibitively expensive in practice as the data volumes grow? We study the above questions on two datasets with thousands of labelled news items, and our results show that: (1) GNNs can achieve comparable or superior performance without any text information to state-of-the-art methods. (2) GNNs trained on a given dataset may perform poorly on new, unseen data, and direct incremental training cannot solve the problem---this issue has not been addressed in the previous work that applies GNNs for fake news detection. In order to solve the problem, we propose a method that achieves balanced performance on both existing and new datasets, by using techniques from continual learning to train GNNs incrementally.
\end{abstract}

%
%

%

\maketitle

\section{Introduction}\label{sec:intro}
While social media has facilitated the timely delivery of various types of information around the world, a consequence is that news is emerging at an unprecedented rate, making it increasingly difficult to fact-check. A series of incidents over recent years have demonstrated the significant damage fake news can cause to society. Therefore, how to automatically and accurately identify fake news before it is widespread has become an urgent challenge for research. Here we use the definition in ~\cite{zhou_fake_2018}: \textit{fake news is intentionally and verifiably false news published by a news outlet}---similar definitions have also been used in previous studies on fake news detection~\cite{monti_fake_2019,shu_defend_2019,shu_fake_2017,ruchansky_csi_2017}.

In our work, we focus on a propagation-based approach for fake news detection. In other words, we use the propagation pattern of news on social media, \eg tweets and retweets of news on Twitter, to determine whether it is fake or not. The feasibility of this approach builds on (1) empirical evidence that fake news and real news spread differently online~\cite{vosoughi_spread_2018}; and (2) the latest development in graph neural networks (GNNs)~\cite{bruna_spectral_2013,niepert_learning_2016,ying_hierarchical_2018,wu_comprehensive_2019} that has enhanced the performance of machine learning models on non-Euclidean data. In addition, as pointed out in~\cite{monti_fake_2019}, whereas content-based approaches require syntactic and semantic analyses, propagation-based approaches are language-agnostic, and can be less vulnerable to adversarial attacks~\cite{szegedy_intriguing_2013,goodfellow_explaining_2014}, where advanced news fabricators carefully manipulate the content in order to bypass detection.

The idea of using propagation patterns to detect fake news has been explored in a number of previous studies~\cite{wu_false_2015,ma_detect_2017,wu_tracing_2018,liu_early_2018,zhou_network-based_2019,shu_hierarchical_2019}, where different types of models have been considered: Wu \etal~\cite{wu_false_2015} use a hybrid Support Vector Machine (SVM), Ma \etal~\cite{ma_detect_2017} use Propagation Tree  Kernel; Wu \etal~\cite{wu_tracing_2018}  incorporate Long Short-Term Memory (LSTM) cells into the Recurrent Neural Network (RNN) model; Liu \etal~\cite{liu_early_2018} use both RNNs and Convolutional Neural Networks (CNNs); Shu \etal~\cite{shu_hierarchical_2019} and Zhou \etal~\cite{zhou_network-based_2019} propose different types of features and compare multiple commonly used machine learning models. The most relevant works include~\cite{monti_fake_2019,lu_gcan_2020,bian_rumor_2020}, which also apply GNNs to study propagation patterns. However, in addition to selecting a different GNN algorithm specifically designed for graph classification (refer to Section~\ref{sec:background} for further explanation), our work mainly focuses on the following questions:

\begin{itemize}[wide=0pt]
    \item \textbf{Question 1: Without relying on any text information, e.g., tweet content, replies and user descriptions, how accurately can GNNs identify fake news?} It is demonstrated in Section~\ref{sec:approach} that even though our model is limited to a restricted set of non-textual features obtained from user profiles and timeline tweets, 
    GNNs can be trained on propagation patterns and these features to achieve comparable or superior performance to state-of-the-art methods that require sophisticated analyses on tweet content, user replies, etc. We argue that the limited set of features can further enhance the security of our models against adversarial attacks, as previous work has shown that high dimensionality facilitates the generation of adversarial samples, resulting in an increased attack surface~\cite{wang_theoretical_2016}.
    \item \textbf{Question 2: How to deal with new, unseen data?} The above question is only concerned with the performance of GNNs on a single dataset. However, a trained model may face vastly different data in practice, and it is important to further investigate how models perform in this scenario. Specifically, we find that GNNs trained on a given dataset may perform poorly on another dataset, and direct incremental training cannot solve the problem---this issue has not been discussed in the previous work that uses GNNs for fake news detection. In order to solve the problem, we propose a method that applies techniques from continual learning to train GNNs incrementally, so that they achieve balanced performance on both existing and new datasets. The method avoids re-training the model on the entire data from scratch---new data always exist, and this becomes prohibitively expensive as data volumes grow.
\end{itemize}

The remainder of this paper is organised as follows: 
Section~\ref{sec:background} briefly introduces the background on graph neural networks; 
Section~\ref{sec:approach} describes our content-free, GNNs-based fake news detection algorithm; 
Section~\ref{sec:new_data} investigates how to deal with new, unseen data, and proposes a solution to achieve balanced performance on both existing and new data by applying techniques from continual learning; 
Section~\ref{sec:related} reviews previous work in fake news detection on social media; 
and finally Section~\ref{sec:conc} concludes the paper and offers directions for future work.

\section{Background on Graph Neural Networks}\label{sec:background}
Although deep learning has witnessed tremendous success in a wide range of applications, including image classification, natural language processing and speech recognition, it mostly deals with data in Euclidean space. GNNs~\cite{bruna_spectral_2013,niepert_learning_2016,ying_hierarchical_2018,wu_comprehensive_2019}, by contrast, are designed to process data generated from non-Euclidean domains.

Consider a graph \(G = (A, F)\) with \(n\) vertices/nodes and \(m\) edges, where \(A \in \{0, 1\}^{n\times n}\) is the adjacency matrix. \(A_{i,j}=1\) if there is an edge from node \(i\) to node \(j\), and \(A_{i,j}=0\) otherwise; \(F \in R^{n\times d}\) is the feature matrix, \ie each node has \(d\) features. Given \(A\) and \(F\) as inputs, the output of a GNN, \ie node embeddings, after the \(k^{th}\) step is: \(H^{(k)} = f\left(A, H^{(k-1)}; \theta^{(k)}\right) \in R^{n \times d}\), where \(f\) is the propagation function parameterised by \(\theta\), and \(H^{(0)}\) is initialised by the feature matrix, \ie \(H^{0} = F\).

There have been a number of implementations for the propagation function. A simple form of the function is: \(f\left(A, H^{(k)}\right) = \allowbreak \sigma\left(AH^{(k-1)}W^{(k)}\right)\), where \(\sigma\) is a non-linear activation function, \eg the rectified linear unit (ReLU) function, and \(W^{(k)}\) is the weight matrix for layer \(k\). A popular implementation of the function is~\cite{kipf_semi-supervised_2017}: \(f\left(A, H^{(k)}\right) = \allowbreak \sigma\left(\tilde{D}^{-\frac{1}{2}}\tilde{A}\tilde{D}^{-\frac{1}{2}}H^{(k-1)}W^{(k)}\right)\), where \(\tilde{A} = A + I\), \(\tilde{D} = \sum_{j}\tilde{A}_{ij}\). Please refer to~\cite{wu_comprehensive_2019} for more choices of the function.

GNNs can perform node regression, node classification, link prediction, edge classification or graph classification depending on the requirements. In our work, since the goal is to label the propagation pattern of each item of news, which is a graph, we choose the algorithm of DiffPool~\cite{ying_hierarchical_2018} that is specifically designed for graph classification. DiffPool extends any existing GNN model by further considering the structural information of graphs. At each layer DiffPool takes the original output \(H^{(k)}\) and the adjacency matrix \(A\), and learns a coarsened graph of \(n^{\prime} < n\) nodes, with the adjacency matrix \(A^{\prime} \in R^{n^{\prime} \times n^{\prime}}\) and the node embeddings \(H^{\prime} \in R^{n^{\prime} \times d}\).

\section{Propagation-based Fake News Detection}\label{sec:approach}
As mentioned in the introduction, we use the definition in ~\cite{zhou_fake_2018} that fake news is intentionally and verifiably false news published by a news outlet. Once an item of news is published, it may be tweeted by multiple users. We call these tweets that directly reference the news URL \textit{root} tweets. Each of them and their retweets form a separate cascade~\cite{vosoughi_spread_2018}, and all the cascades form the propagation pattern of an item of news. The purpose of this work is to determine the validity of an item of news using its propagation pattern.

Formally, we define the propagation-based fake news detection problem as follows: given a set of labeled graphs \(\mathcal{D} = \{\left(G_{1}, y_{1}\right),\allowbreak \left(G_{2}, y_{2}\right),\allowbreak ...,\allowbreak \left(G_{i}, y_{i}\right), ...\}\), where \(G_{i} \in \mathcal{G}\) is the propagation pattern for news \(i\), and \(y_{i} \in \mathcal{Y} = \{0\ \text{(Real)},\allowbreak 1\ \text{(Fake)}\}\) is the label of graph \(G_{i}\), the goal is to learn a mapping \(g: \mathcal{G} \to \mathcal{Y}\) that labels each graph.

In the remainder of this section, we first explain how we generate a graph in Section~\ref{sec:approach-data}, \ie the adjacency matrix and the feature matrix, and present the experimental results to verify the effectiveness of the GNN-based detection algorithm in Section~\ref{sec:approach-exp}.

\begin{figure}[ht]
    \centering
    \includegraphics[width=0.54\columnwidth]{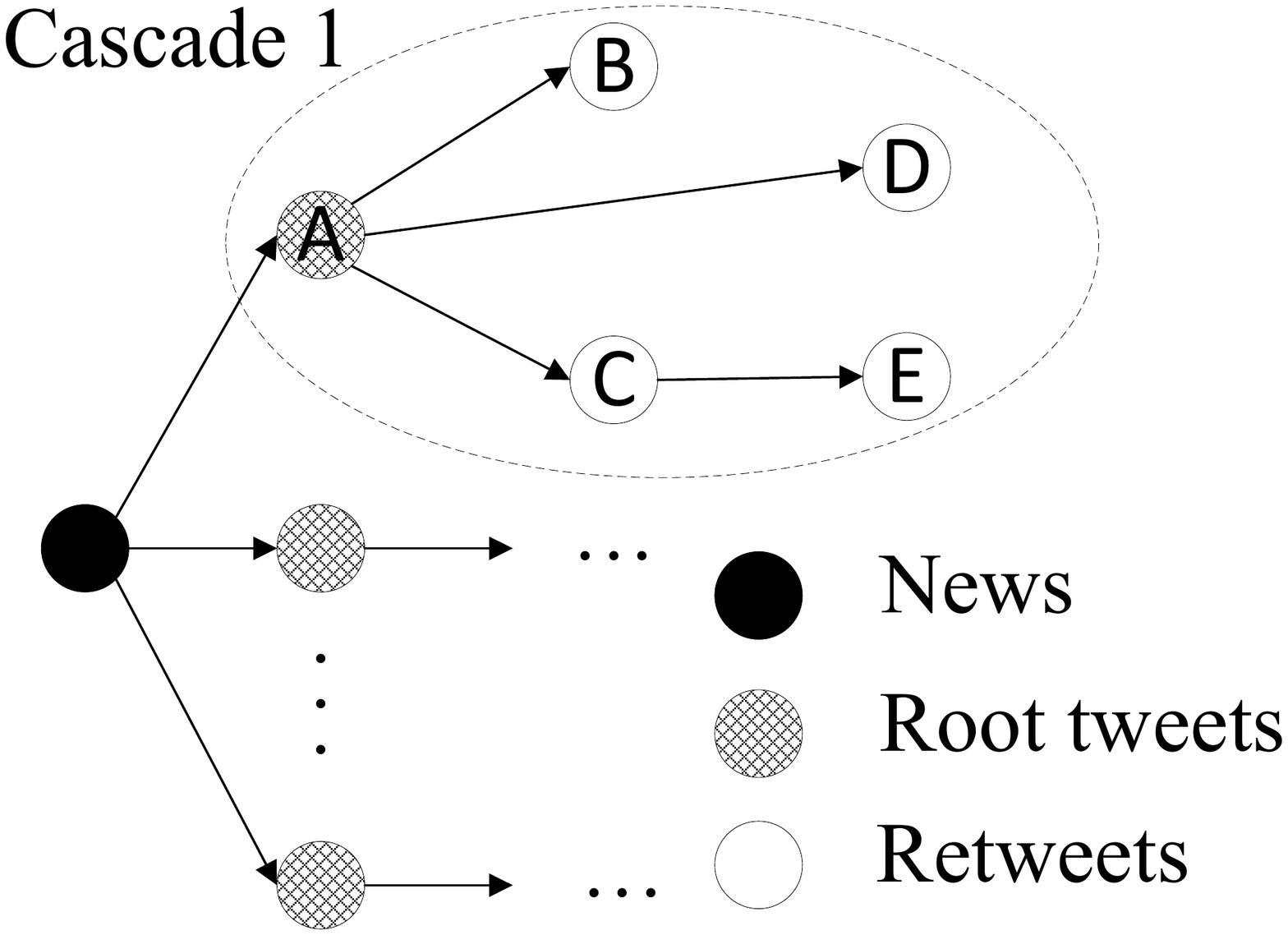}
    \caption{An illustration of the graph for each item of news.}
    \label{fig_graph}
\end{figure}

\begin{figure*}[ht!]
\centering
\begin{subfigure}{.95\columnwidth}
  \centering
  \includegraphics[width=\columnwidth]{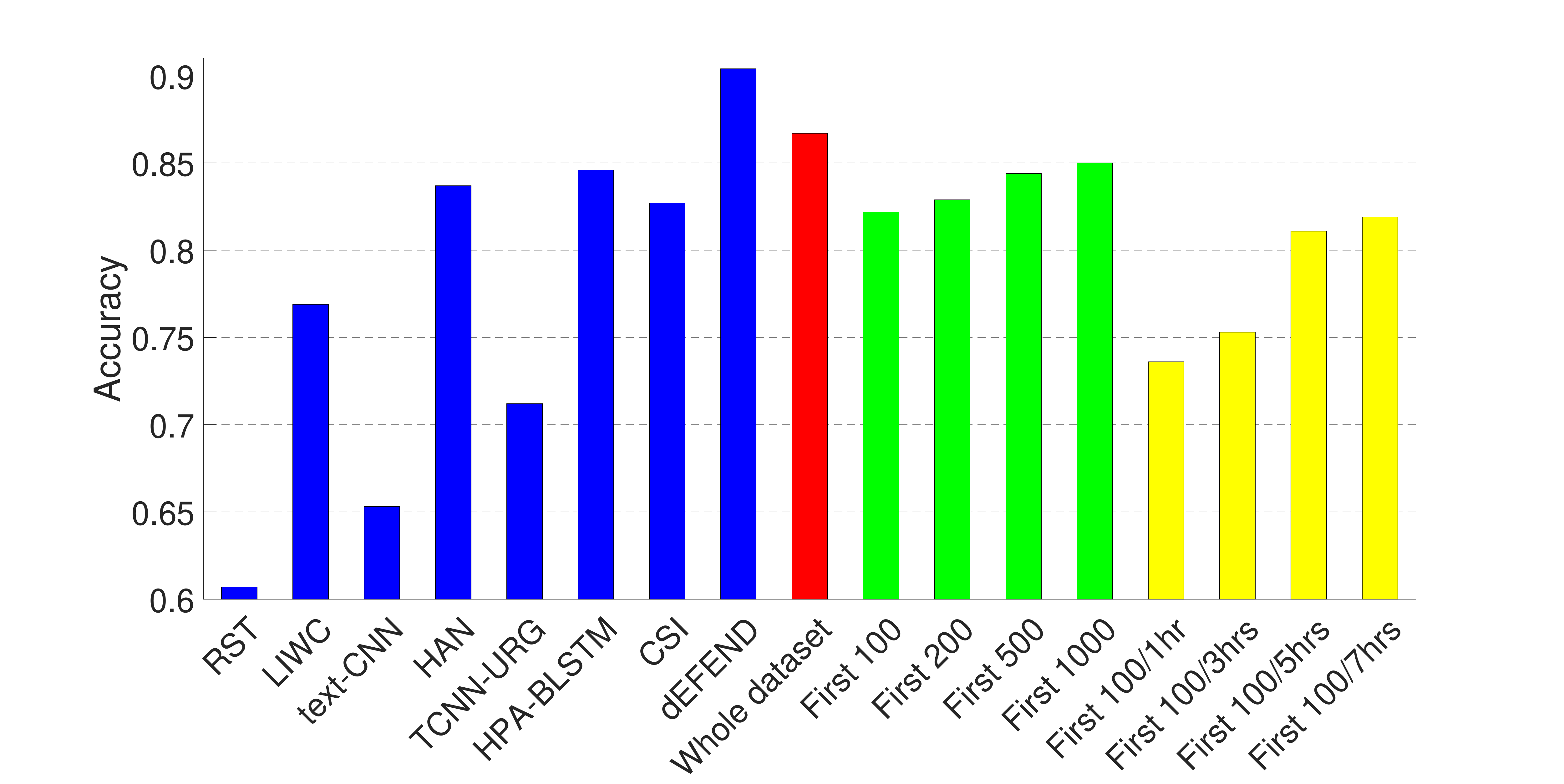}
  \caption{Accuracy}
  \label{figure_plt_complete_acc}
\end{subfigure}
\begin{subfigure}{.95\columnwidth}
  \centering
  \includegraphics[width=\columnwidth]{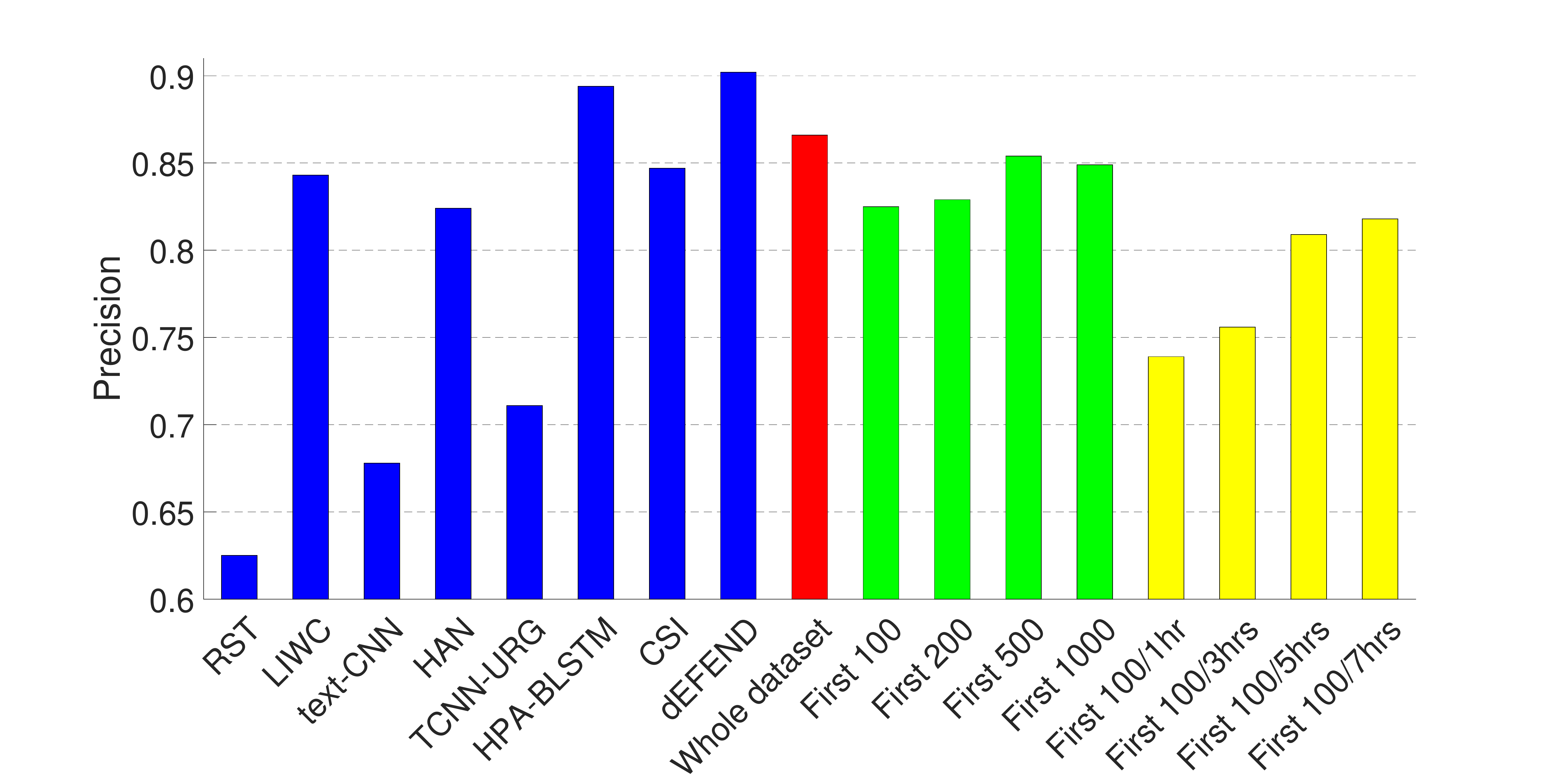}
  \caption{Precision}
  \label{figure_plt_complete_pre}
\end{subfigure}
\begin{subfigure}{.95\columnwidth}
  \centering
  \includegraphics[width=\columnwidth]{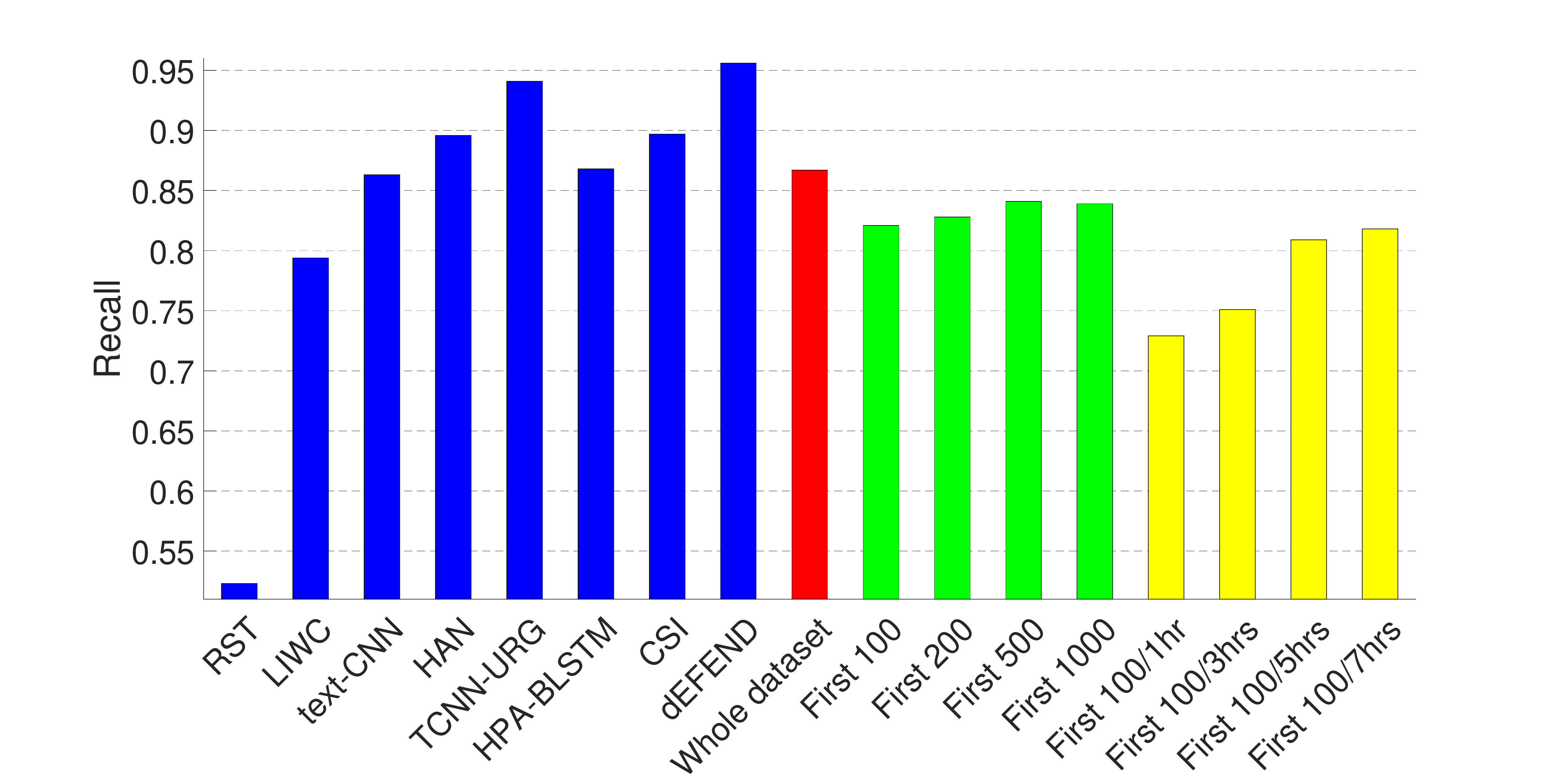}
  \caption{Recall}
  \label{figure_plt_complete_rec}
\end{subfigure}
\begin{subfigure}{.95\columnwidth}
  \centering
  \includegraphics[width=\columnwidth]{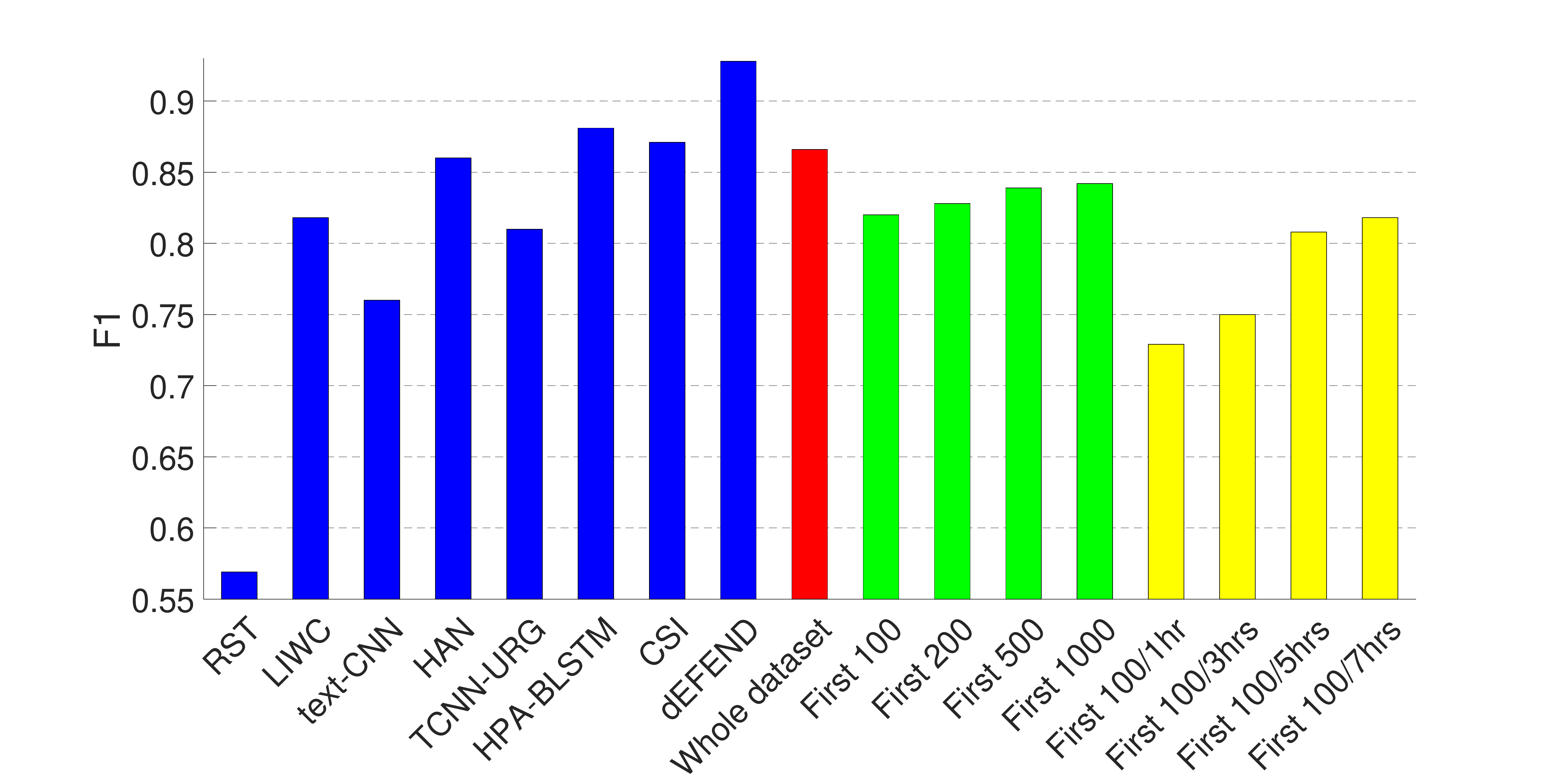}
  \caption{F1}
  \label{figure_plt_complete_f1}
\end{subfigure}
\caption{Performance comparison on the dataset of PolitiFact. The first eight bars correspond to the results of eight fake news detection algorithms as reported in~\cite{shu_defend_2019}, the red bar is the result of our propagation-based method trained on the whole dataset, and the rest are the results of our propagation-based method trained on the clipped datasets for fake news early detection.}
\label{figure_plt_complete}
\end{figure*}

\subsection{Data Generation}\label{sec:approach-data}
In order to generate the news propagation pattern, we use the dataset of FakeNewsNet~\cite{shu_fakenewsnet_2018}, which is especially collected for the purpose of fake news detection. FakeNewsNet contains labelled news from two websites: politifact.com\footnote{https://www.politifact.com/} and gossipcop.com\footnote{https://www.gossipcop.com/}---the news content includes both linguistic and visual information, all the tweets and retweets for each item of news, and the information of the corresponding Twitter users (refer to~\cite{shu_fakenewsnet_2018} for more details).

\textbf{Adjacency matrix.} As illustrated in Fig.~\ref{fig_graph}, each item of news is represented as a graph, where a node refers to a tweet (including the corresponding user)---either the root tweet that references the news or its retweets. A special case is that an extra node representing the news is added to connect all cascades together. All the feature values for this node are set to zero. Edges here represent information flow, \ie how the news transfers from one person to another. However, Twitter APIs do not provide the immediate source of a retweet, \eg in Cascade 1 of Fig.~\ref{fig_graph}, Twitter APIs only show that \(B, C, D,\) and \(E\) are retweets of \(A\), but \(E\) is actually a retweet of \(C\). To solve this problem, within each cascade we first sort the tweets by their timestamps, and then search for the source of a retweet from all the tweets that are published earlier. Specifically, there is an edge from node \(i\) to node \(j\) \footnote{Node \(i\) is published before node \(j\), and the information goes from user \(i\) to user \(j\).} if:

\begin{itemize}
    \item The user of node \(i\) mentions the user of node \(j\) in the tweet, \eg user \(i\) retweets a news item and also recommends it to user \(j\) via mentioning;
    \item Tweet \(i\) is public and tweet \(j\) is posted within a certain period of time after tweet \(i\). We have tested different time limits from one hour to ten hours.
\end{itemize}

Note that edges only exist between nodes within the same cascade. We have also compared the difference by further considering the follower and following relations, but our results demonstrate that there is no significant improvement. In addition, since Twitter applies a much stricter rate limit on corresponding APIs, these types of information may not be available in real time, especially if a number of news items need to be validated at the same time and within a detection deadline. More details on this are given in the next subsection.

\begin{figure*}[ht!]
\centering
\begin{subfigure}{.95\columnwidth}
  \centering
  \includegraphics[width=\columnwidth]{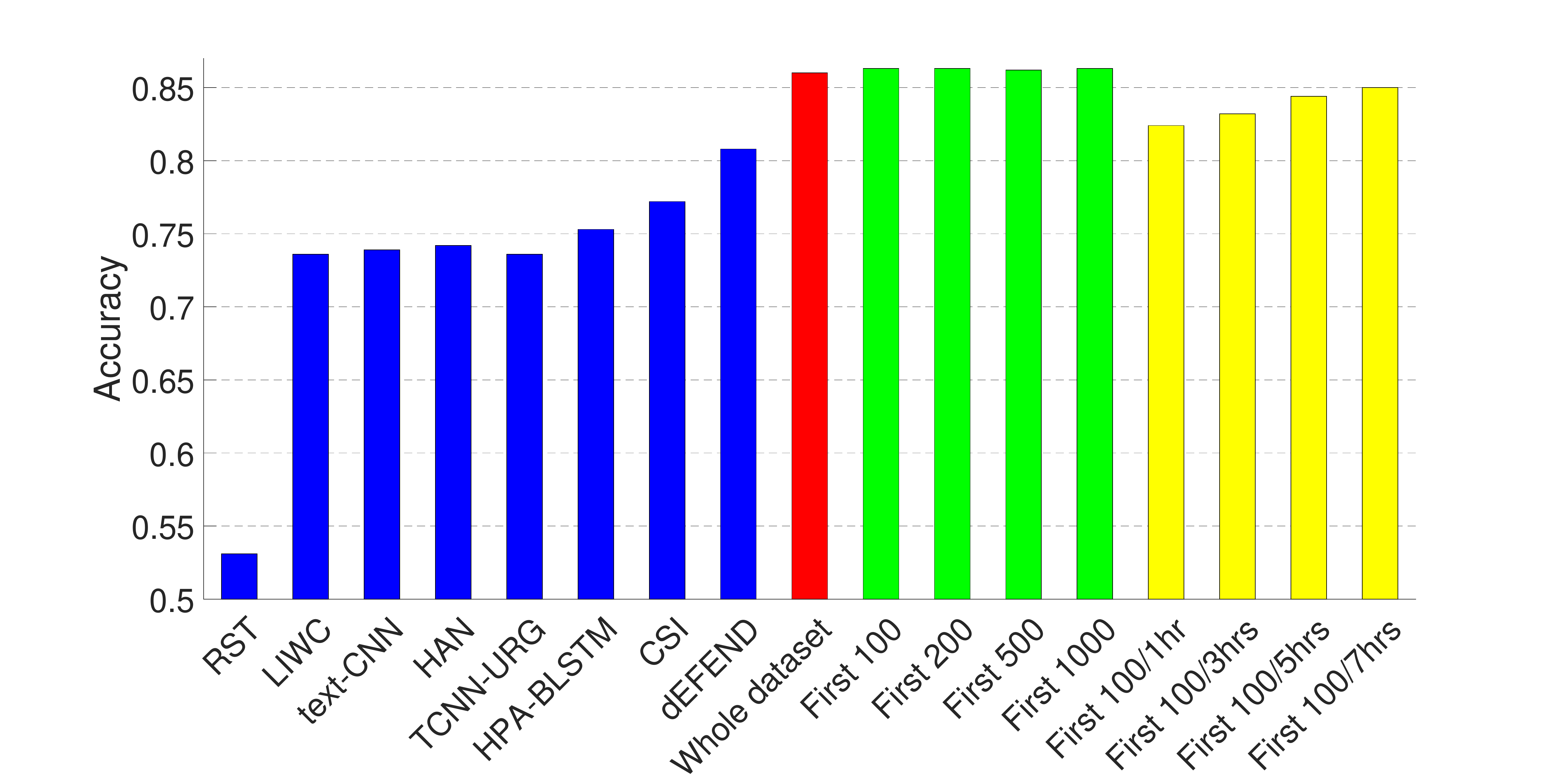}
  \caption{Accuracy}
  \label{figure_gsp_complete_acc}
\end{subfigure}
\begin{subfigure}{.95\columnwidth}
  \centering
  \includegraphics[width=\columnwidth]{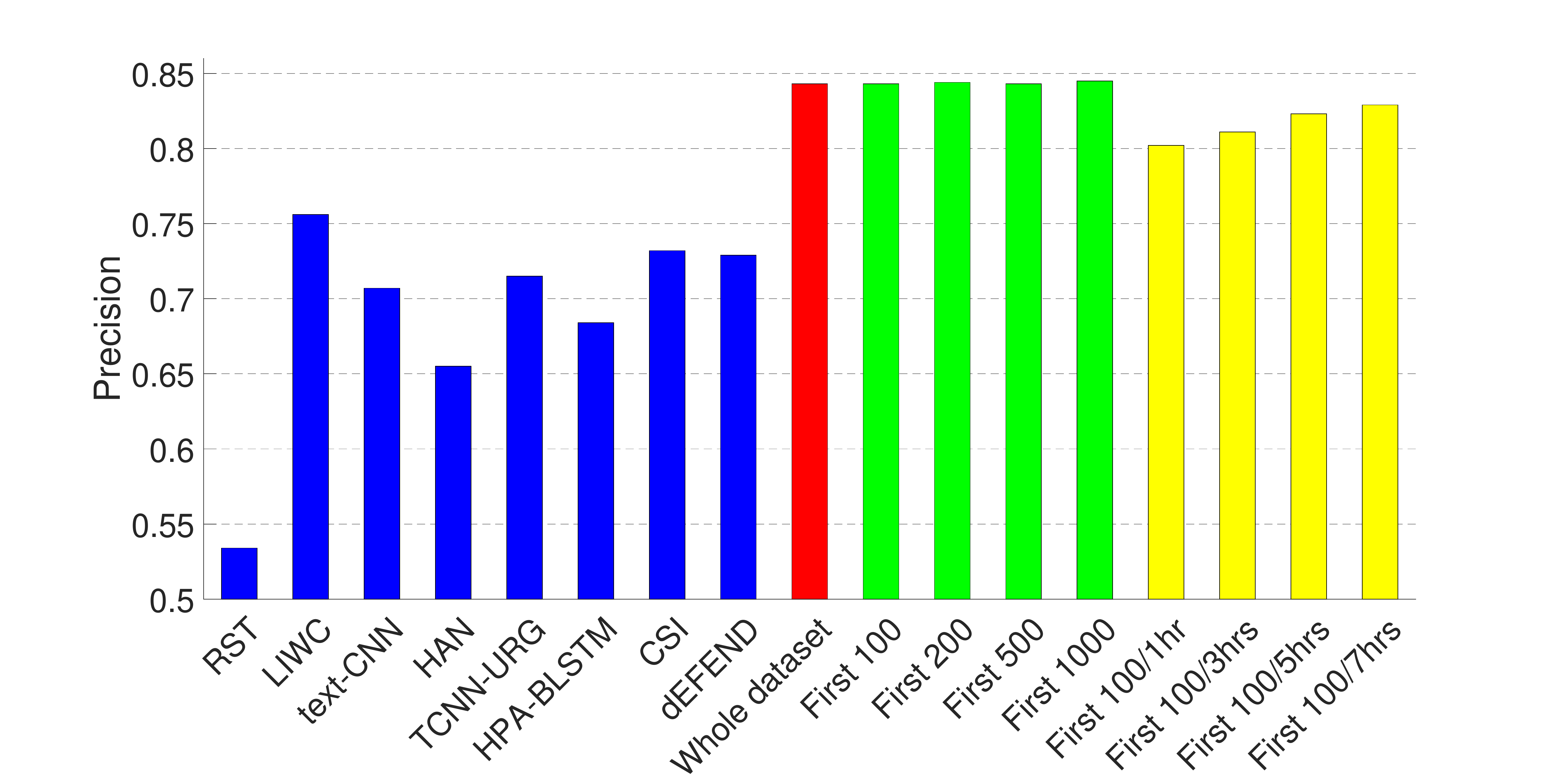}
  \caption{Precision}
  \label{figure_gsp_complete_pre}
\end{subfigure}
\begin{subfigure}{.95\columnwidth}
  \centering
  \includegraphics[width=\columnwidth]{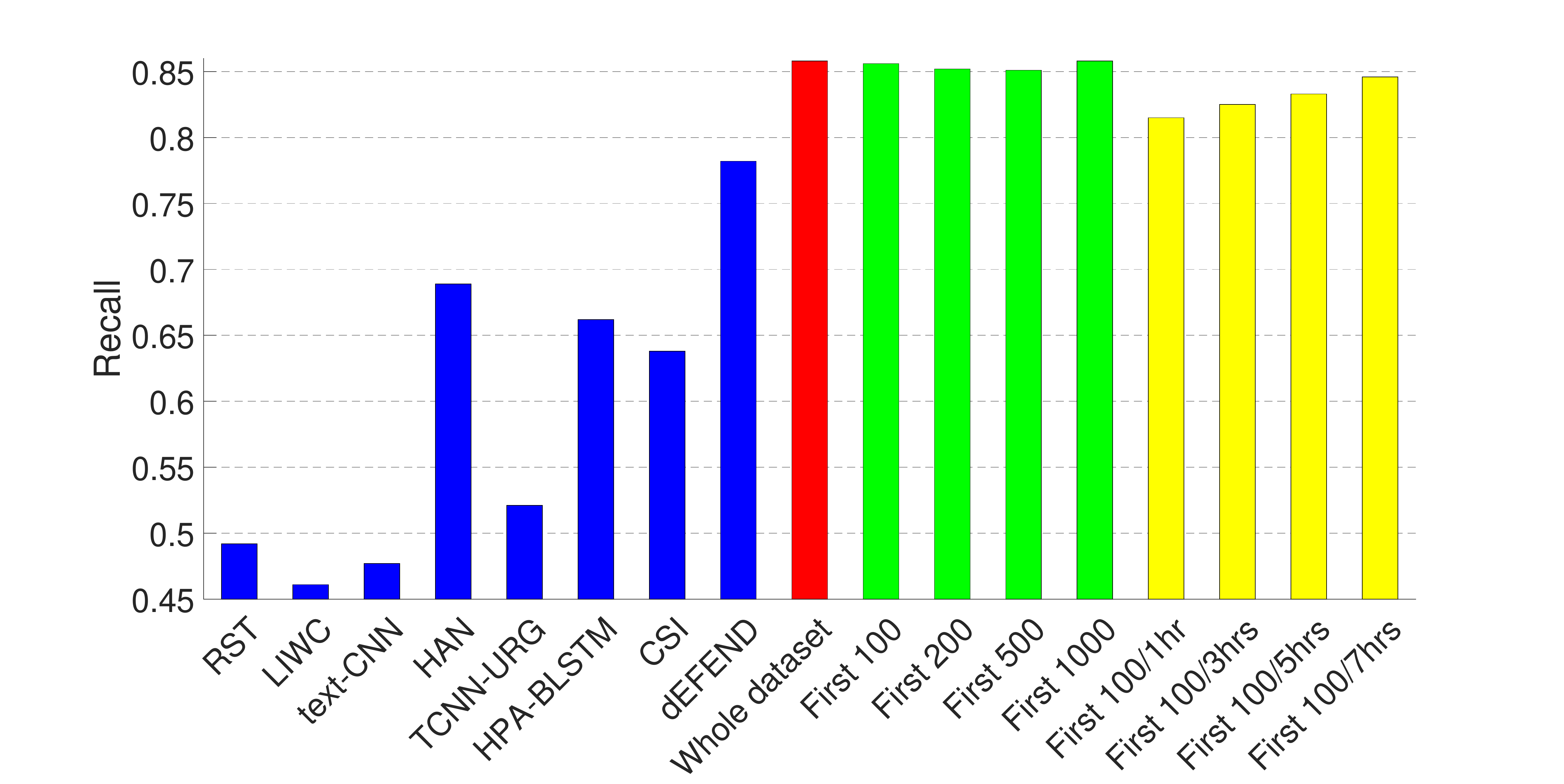}
  \caption{Recall}
  \label{figure_gsp_complete_rec}
\end{subfigure}
\begin{subfigure}{.95\columnwidth}
  \centering
  \includegraphics[width=\columnwidth]{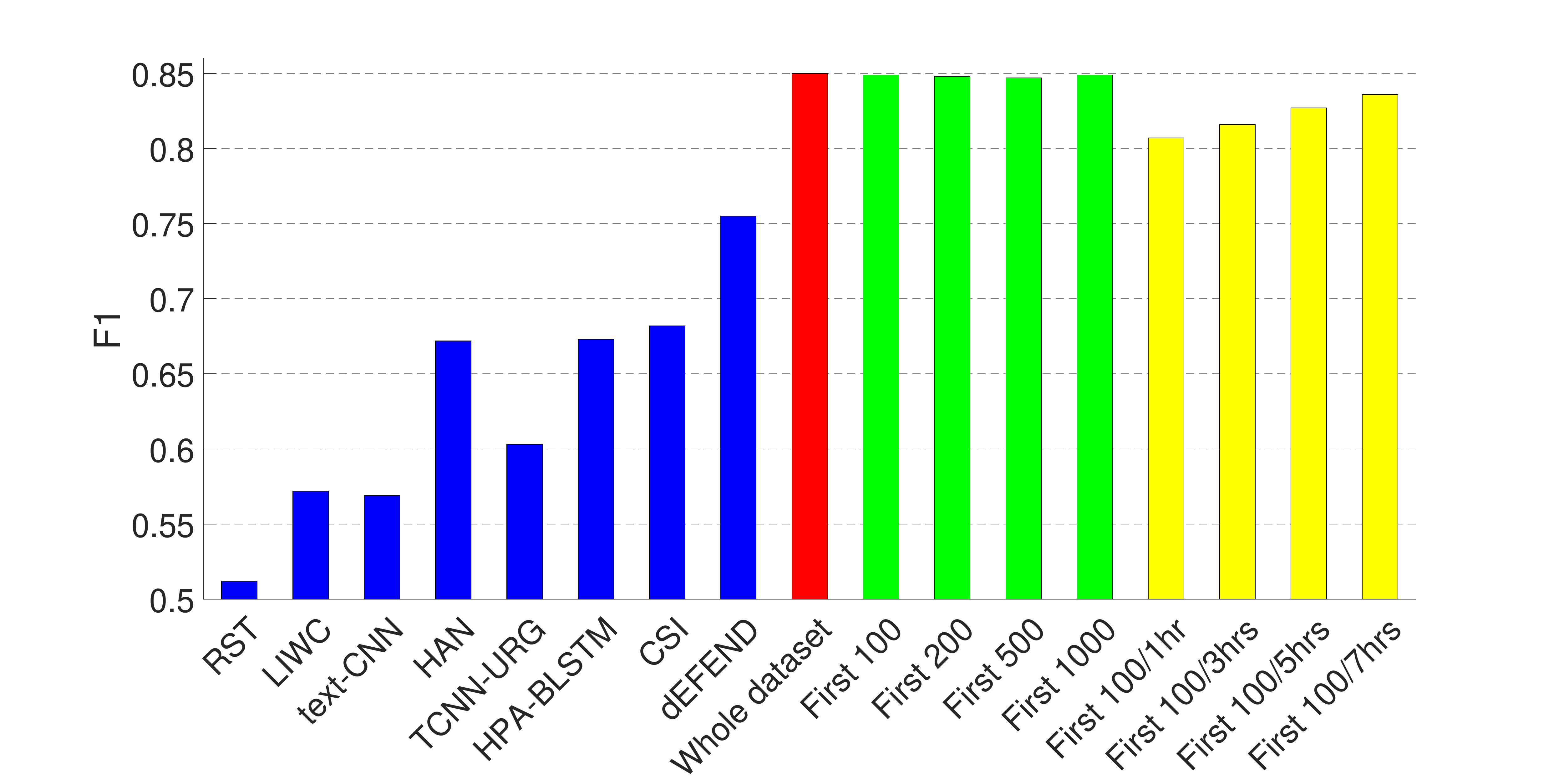}
  \caption{F1}
  \label{figure_gsp_complete_f1}
\end{subfigure}
\caption{Performance comparison on the dataset of GossipCop. The first eight bars correspond to the results of eight fake news detection algorithms as reported in~\cite{shu_defend_2019}, the red bar is the result of our propagation-based method trained on the whole dataset, and the rest are the results of our propagation-based method trained on the clipped datasets for early detection.}
\label{figure_gsp_complete}
\end{figure*}

\textbf{Feature matrix.} As mentioned earlier we do not rely on any textual information in this work, including tweet content, user reply or user description, and only choose the following information from user profiles as the features for each node: 

\begin{itemize}
    \item Whether the user is verified;
    \item The timestamp when the user was created, encoded as the number of months since March 2006---the time when Twitter was founded;
    \item The number of followers;
    \item The number of friends;
    \item The number of lists;
    \item The number of favourites;
    \item The number of statuses;
    \item The timestamp of the tweet, encoded as the number of seconds since the first tweet references the news is posted.
\end{itemize}

Another important reason why we choose the above features is that they are most accessible---they are directly available within the tweet object, which is preferable for online detection.

In addition, based on the hypothesis that less credible users are more likely to form larger clusters than more credible users~\cite{shu_leveraging_2020}, we extract another set of features from user timeline tweets to check if they can further improve the performance of our model. Specifically, we collect the timeline tweets for all the users in the propagation pattern of a news item (a maximum of 200 tweets are collected per user), and construct another graph, where each node represents a user, while an edge exists from node \(i\) to node \(j\) if user \(i\) mentions user \(j\), and the weight of the edge is the number of times that user \(j\) is mentioned by user \(i\). Finally, after the graph is built we calculate the following features for each node \(i\):

\begin{itemize}
    \item The in-degree, \ie the number of users that have mentioned user \(i\);
    \item The out-degree, \ie the number of users that have been mentioned by user \(i\)
    \item The weighted in-degree, \ie the number of times that user \(i\) have been mentioned;
    \item The weighted out-degree, \ie the number of times that user \(i\) have mentioned others;
    \item The number of hop-2 in-neighbours;
    \item The number of hop-2 out-neighbours;
    \item The number of collected timeline tweets.
\end{itemize}

The rationale of studying these features is that less credible users are more likely to collaborate with each other, and such behaviour can be captured by the above features.

In our experiment, we first train models only using the features from user profiles, and then compare the difference with (1) training models on the features from timeline tweets, and (2) training models on a combination of both sets of features.

\subsection{Experimental Verification}\label{sec:approach-exp}
Using the method introduced in the previous subsection to generate the graphs (the adjacency and feature matrices), we test multiple DiffPool models with a range of different architectures: 2-4 pooling layers, 16-128 hidden dimensions and 16-128 embedding dimensions (the chosen hyper-parameters under different settings are given later in this section). As recommended by the authors in~\cite{ying_hierarchical_2018}, we use DiffPool built on top of GraphSage~\cite{hamilton_inductive_2017}.

In order to make our results comparable with those reported in~\cite{shu_defend_2019} (as they also tested fake news detection algorithms on the same dataset), we follow the same procedure to train and test the GNNs: randomly choose 75\% of the news as the training data while keeping the rest as the test data, and the final result is the average performance over five repeats. In addition, the model is evaluated with the following commonly used metrics: accuracy, precision, recall and F1 score. The main reason why we do not use the same split of training and test data across all experiments is that an algorithm may perform extremely well on one split of data but rather poorly on another. Therefore, all algorithms are tested on multiple random splits of data, so that the obtained results are closer to their real performance.

\begin{table*}[ht!]
    \centering
    \caption{Performance comparison of models trained (1) with/without non-textual features extracted from user timeline tweets, and (2) with/without considering the follower/following relation.}
    \begin{tabular}{|c|c|c|c|c|c|c|c|}
        \hline
        \multirow{3}{*}{Dataset} & \multirow{3}{*}{Metric} & \multicolumn{2}{c|}{User profile features only} & \multicolumn{2}{c|}{Timeline tweets features only} &
        \multicolumn{2}{c|}{Combined} \\
        \cline{3-8}
        & & Without follower/ & With follower/ & Without follower/ & With follower/ & Without follower/ & With follower/ \\
        & & following & following & following & following & following & following \\
        \hline
        \multirow{4}{*}{PolitiFact} & Acc & 0.811 & 0.805 & 0.699 & 0.696 & 0.792 & 0.803 \\
        \cline{2-8}
        & Pre & 0.809 & 0.806 & 0.700 & 0.694 & 0.792 & 0.806 \\
        \cline{2-8}
        & Rec & 0.809 & 0.801 & 0.695 & 0.695 & 0.792 & 0.801 \\
        \cline{2-8}
        & F1 & 0.808 & 0.801 & 0.693 & 0.691 & 0.791 & 0.801 \\
        \hline
        \multirow{4}{*}{GossipCop} & Acc & 0.844 & 0.841 & 0.853 & 0.853 & 0.849 & 0.841 \\
        \cline{2-8}
        & Pre & 0.823 & 0.821 & 0.834 & 0.831 & 0.829 & 0.820 \\
        \cline{2-8}
        & Rec & 0.833 & 0.834 & 0.852 & 0.846 & 0.840 & 0.831 \\
        \cline{2-8}
        & F1 & 0.827 & 0.826 & 0.841 & 0.837 & 0.833 & 0.825 \\
        \hline
    \end{tabular}
    \label{table_performance_timeline_follow}
\end{table*}

\subsubsection{Training on the Complete Dataset}
We first train GNNs on the whole dataset of PolitiFact/GossipCop, using the features from user profile only, and without considering the follower/following relation. After testing a range of hyper-parameters, we find that a four-layer GNN with 64 hidden dimensions and 64 embedding dimensions works best for the dataset of PolitiFact, while for GossipCop the number of layers should decrease to two as there are significantly more news items in this dataset.

The experimental results are presented in Figs.~\ref{figure_plt_complete} and~\ref{figure_gsp_complete}, where (1) The first eight bars correspond to the results of eight fake news detection algorithms as reported in~\cite{shu_defend_2019} on the same dataset---RST~\cite{rubin_towards_2015}, LIWC~\cite{pennebaker_development_2015}, HAN~\cite{yang_hierarchical_2016}, text-CNN~\cite{kim_convolutional_2014}, TCNN-URG~\cite{qian_neural_2018}, HPA-BLSTM~\cite{guo_rumor_2018}, CSI~\cite{ruchansky_csi_2017} and dEFEND~\cite{shu_defend_2019}. Note that all of these methods require analysis on textual information, \eg tweet content and user replies. (2) The ninth bar in red is the result of our propagation-based method trained on the whole dataset.

As can be seen from the figures, by only relying on the limited set of non-textual features as introduced in Section~\ref{sec:approach-data}, our model can achieve comparable performance on the dataset of PolitiFact, and the best result on the dataset of GossipCop.

\subsubsection{Training on the Partial Dataset for Early Detection}\label{sec:early_detection}
It is critical to detect fake news at an early stage before it becomes widespread, since the wider fake news spreads, the more likely people would trust it~\cite{boehm_validity_1994}, and it is difficult to correct people's perception towards an issue, even if the previous impression is inaccurate~\cite{keersmaecker_fake_2017}.

Therefore, we train GNNs on the clipped dataset that contains for each news item (1) the first 100, 200, 500, 1000 tweets (green bars in Figs.~\ref{figure_plt_complete} and~\ref{figure_gsp_complete}); and (2) the first 100 tweets or tweets from the first one, three, five or seven hours, whichever is smaller (yellow bars in Figs.~\ref{figure_plt_complete} and~\ref{figure_gsp_complete}). The hyper-parameters here are the same as in the last set of experiments, except that for the clipped GossipCop dataset that contains the first 100 tweets (both with and without the different time limits), the number of pooling layers is three.

The results demonstrate that even with a limited number of tweets per news item, our model can achieve a decent performance, especially on the dataset of GossipCop, which is likely to be due to the larger size of the dataset.

\subsubsection{Additional Non-textual Features from User Timeline Tweets}
Here we investigate the impact of the set of non-textual features extracted from user timeline tweets as introduced in Section~\ref{sec:approach-data}.

Note that from here forward we focus on models trained on the clipped dataset with the first 100 tweets or the tweets from the first five hours for each item of news, since previous results have demonstrated that models trained on this dataset can achieve reasonably close performance to models trained on the complete dataset, and more importantly it is crucial to detect fake news items before they become widespread.

The results in Table~\ref{table_performance_timeline_follow} (the third, fifth and seventh columns) show that models trained on a combination of the two sets of features do not show obvious improvement over performance, although for the dataset of GossipCop, models trained on the features from timeline tweets alone perform equally well with models previously obtained in Section~\ref{sec:early_detection}.

\subsubsection{Further Considering Follower and Following Relations}
Previously when constructing the adjacency matrix, we have not considered the follower and following relations between Twitter users. In this subsection, we examine whether the results can be further improved by including these types of information, \ie an edge is added from node \(i\) to node \(j\) if user\(j\) follows user \(i\).

Table~\ref{table_performance_timeline_follow} suggests that there is not any significant difference with and without considering the follower/following relation, when the model is trained on the features either from user profiles, timeline tweets or both. Therefore, the relation is not included in our model.

\textbf{Model efficiency.} When training and testing our models, we also find that GNNs converge very quickly---most of the time it only takes dozens of epochs for the model to reach similar performance to the final model in terms of the four metrics, while each epoch lasts from only a couple of seconds to several minutes, depending on the different model structures and sizes of the datasets.

All these results provide strong support for applying GNNs in propagation-based fake news detection.

\section{Dealing with New Data}\label{sec:new_data}
While the above results demonstrate the effectiveness of our proposed method on a single dataset, this section further studies the model performance on new data.

\begin{figure}[ht!]
\centering
\includegraphics[width=0.85\columnwidth]{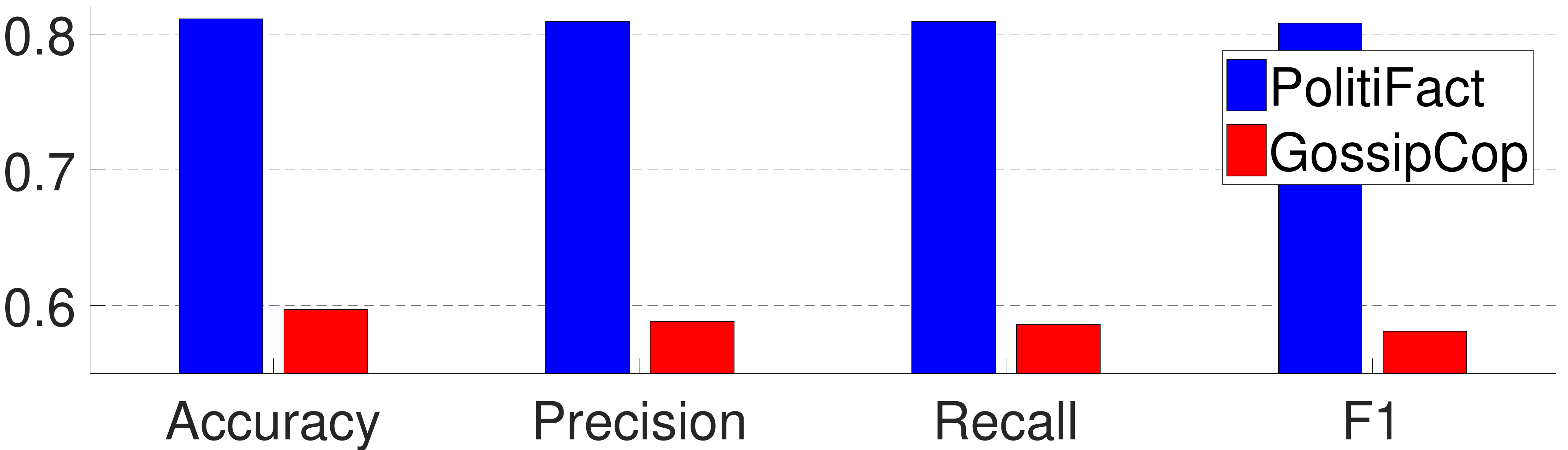}
\caption{Models trained on the clipped dataset of PolitiFact perform poorly on the dataset of GossipCop.}
\label{figure_plt_complete_trained}
\end{figure}

Let one dataset, \eg PolitiFact, represent the existing data that our model has been trained on, and the other dataset, \eg GossipCop, represent the unknown data that our model will face in the future, we find that models trained on PolitiFact do not perform well on GossipCop (Fig.~\ref{figure_plt_complete_trained}), and vice versa (the figure for this case is omitted due to similarity). 

An examination of the graphs reveals that the graphs generated from PolitiFact and GossipCop are vastly different, in terms of the numbers of nodes and edges, which explains the reason for the observed behavior.

\textbf{Why not directly train on both datasets?} A natural thought is to re-train the model on both datasets, but this may not be feasible, or at least not ideal in practice: there will always be new data that our model has not seen before, and it does not make sense to re-train the model from scratch on the entire data every time a new dataset is obtained, especially since as the data size grows, this can become prohibitively expensive. In the remainder of this section, we address the issue of dealing with new, unseen data.

\subsection{Incremental Training}\label{sec:incremental}
\begin{figure}[t!]
\centering
\includegraphics[width=0.85\columnwidth]{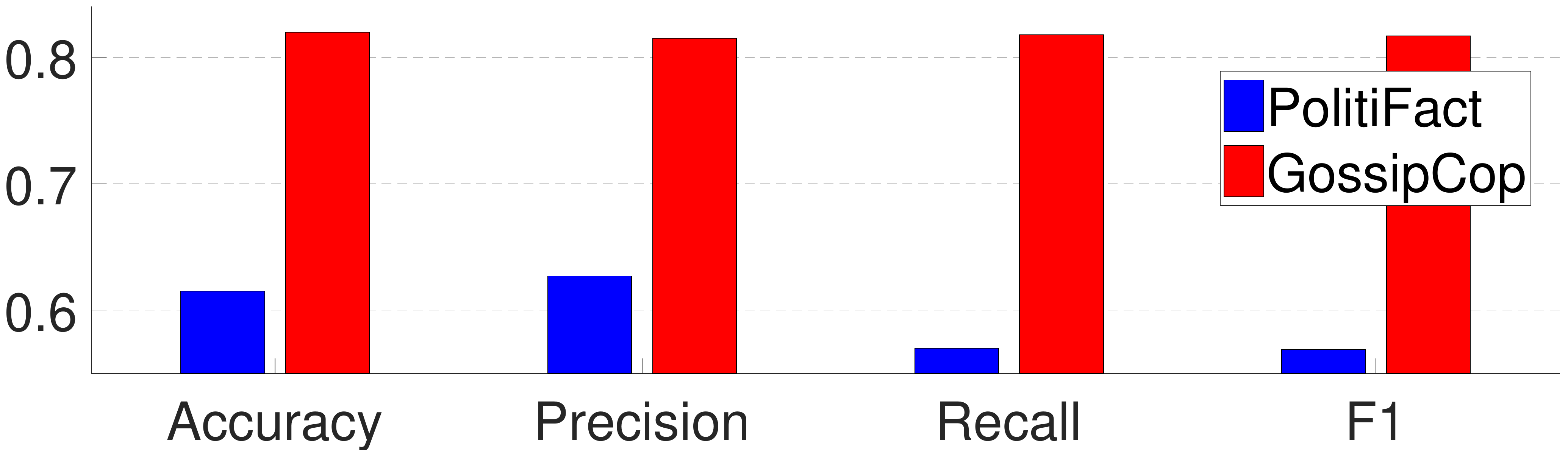}
\caption{Models first trained on the clipped dataset of PolitiFact and then on GossipCop only perform well on the latter dataset on which it is trained, \ie GossipCop.}
\label{figure_cf_plt_gsp}
\end{figure}

We first test incremental training, \ie further train the model obtained from PolitiFact (or GossipCop) on the other dataset of GossipCop (or PolitiFact). However, as shown in Fig.~\ref{figure_cf_plt_gsp}, then the models only perform well on the latter dataset on which they are trained, while achieving degraded results on the former dataset (the figure for the models first trained on GossipCop and then on PolitiFact is omitted due to similarity). Note that during incremental training, we still randomly choose 75\% of graphs as the training data and the rest as the test data.

This is similar to the problem of catastrophic forgetting~\cite{bower_catastrophic_1989,ratcliff_connectionist_1990,mcclelland_why_1995,french_catastrophic_1999} in the field of continual learning: when a deep neural network is trained to learn a sequence of tasks, it degrades its performance on the former tasks after it learns new tasks, as the new tasks override the weights. 

In our case, each new dataset can be considered as a new task. In the next subsection, we investigate how to solve the problem by applying techniques from continual learning.

\subsection{Continual Learning}\label{sec:continual}
In order to deal with catastrophic forgetting, a number of approaches have been proposed, which can be roughly classified into three types~\cite{parisi_continual_2018}: (1) regularisation-based approaches that add extra constraints to the loss function to prevent the loss of previous knowledge; (2) architecture-based approaches that selectively train a part of the network for each task, and expand the network when necessary for new tasks; (3) dual-memory-based approaches that build on top of complementary learning systems (CLS) theory~\cite{mcclelland_why_1995,kumaran_what_2016}, and replay samples for memory consolidation.

In this paper, we choose the following two popular methods: 
\begin{itemize}
    \item Gradient Episodic Memory (GEM)~\cite{lopez-paz_gradient_2017}---GEM uses episodic memory to store a number of samples from previous tasks, and when learning a new task \(t\), it does not allow the loss over those samples held in memory to increase compared to when the learning of task \(t-1\) is finished;
    \item Elastic Weight Consolidation (EWC)~\cite{kirkpatrick_overcoming_2017}---its loss function consists of a quadratic penalty term on the change of the parameters, in order to prevent drastic updates to those parameters that are important to the old tasks.
\end{itemize}

\begin{figure*}[ht]
\centering
\begin{subfigure}{.33\textwidth}
  \centering
  \includegraphics[width=\textwidth]{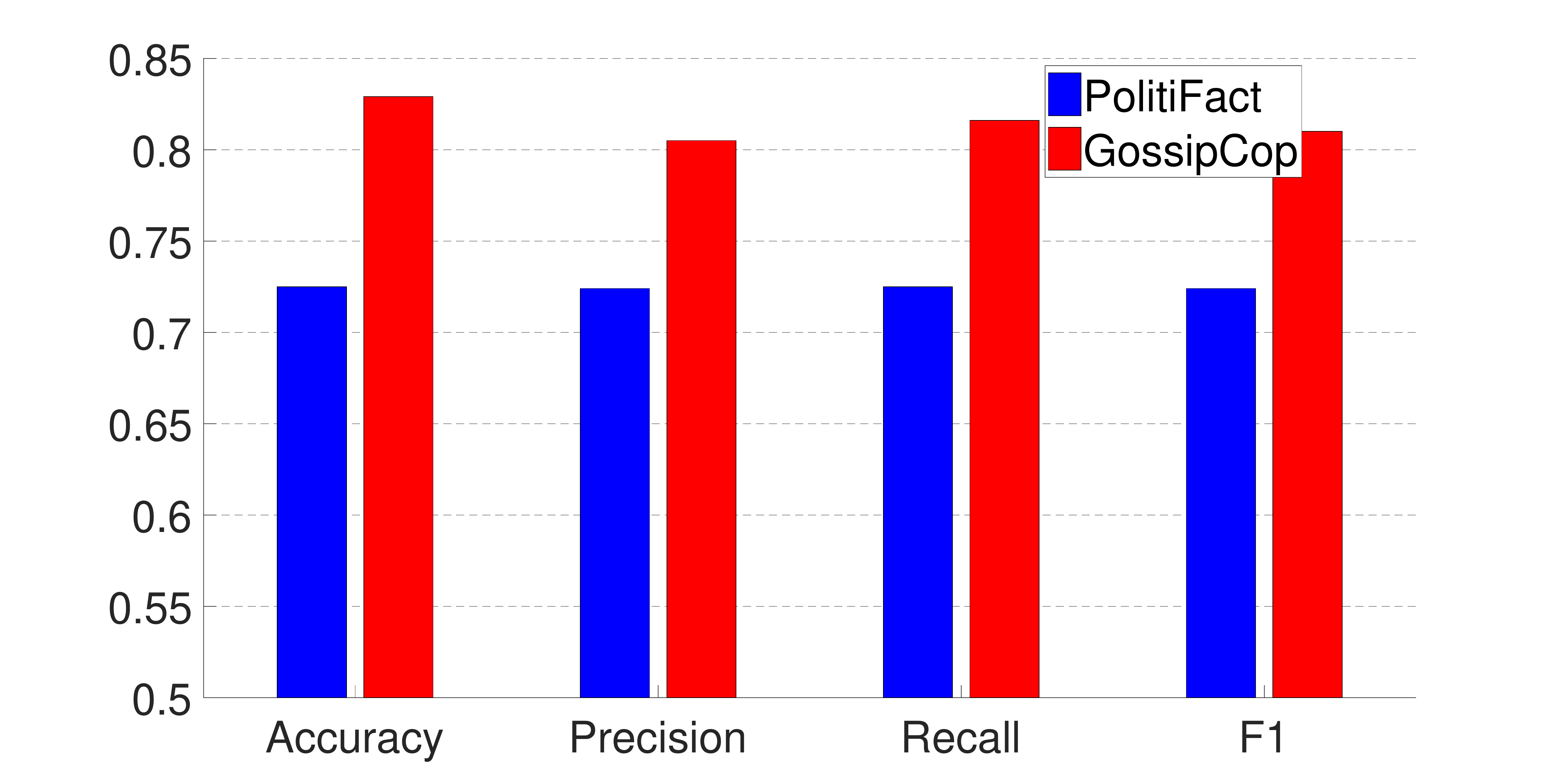}
  \caption{Sample size=100}
  \label{figure_gem_plt_gsp_100}
\end{subfigure}
\begin{subfigure}{.33\textwidth}
  \centering
  \includegraphics[width=\textwidth]{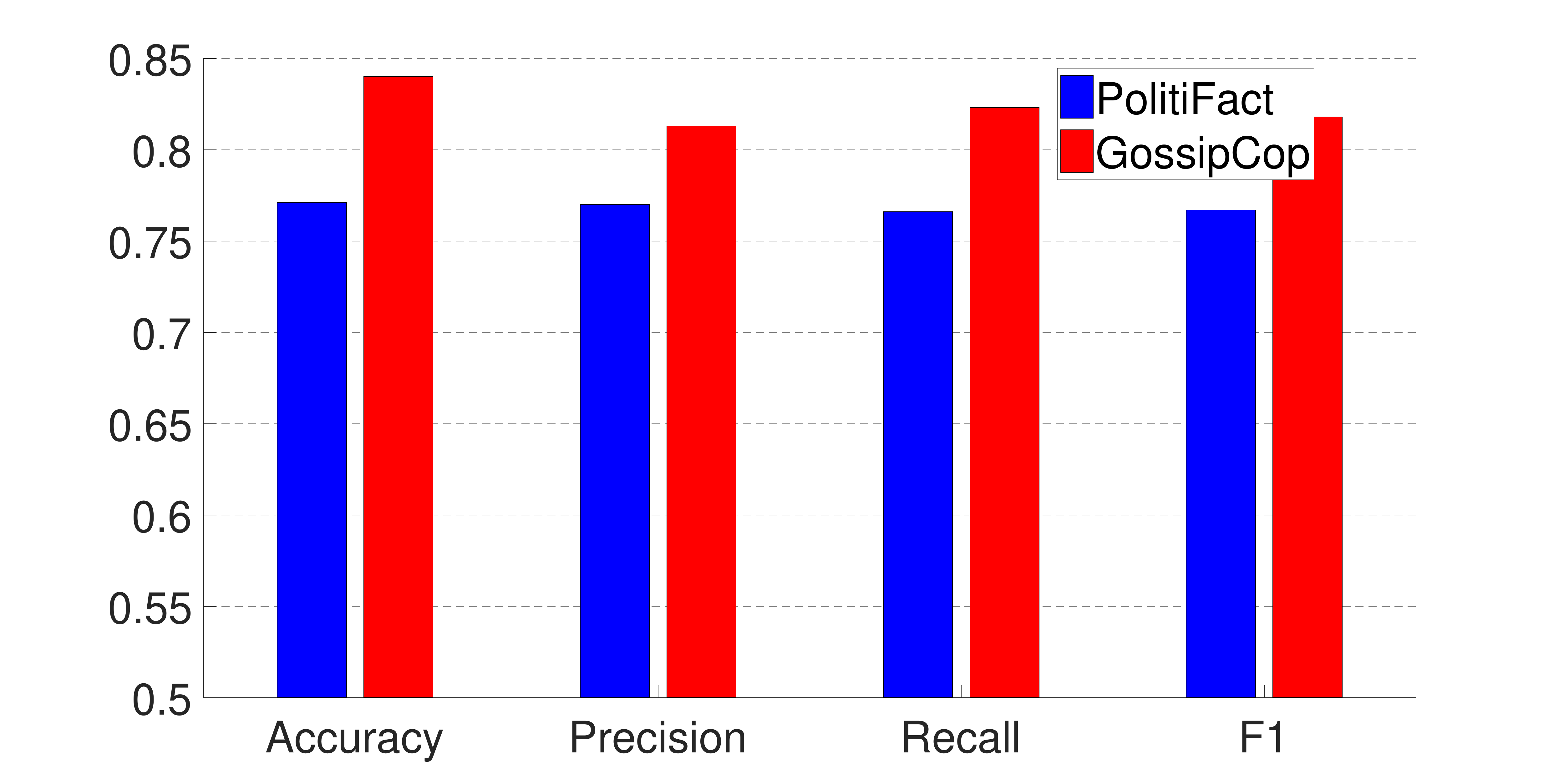}
  \caption{Sample size=200}
  \label{figure_gem_plt_gsp_200}
\end{subfigure}
\begin{subfigure}{.33\textwidth}
  \centering
  \includegraphics[width=\textwidth]{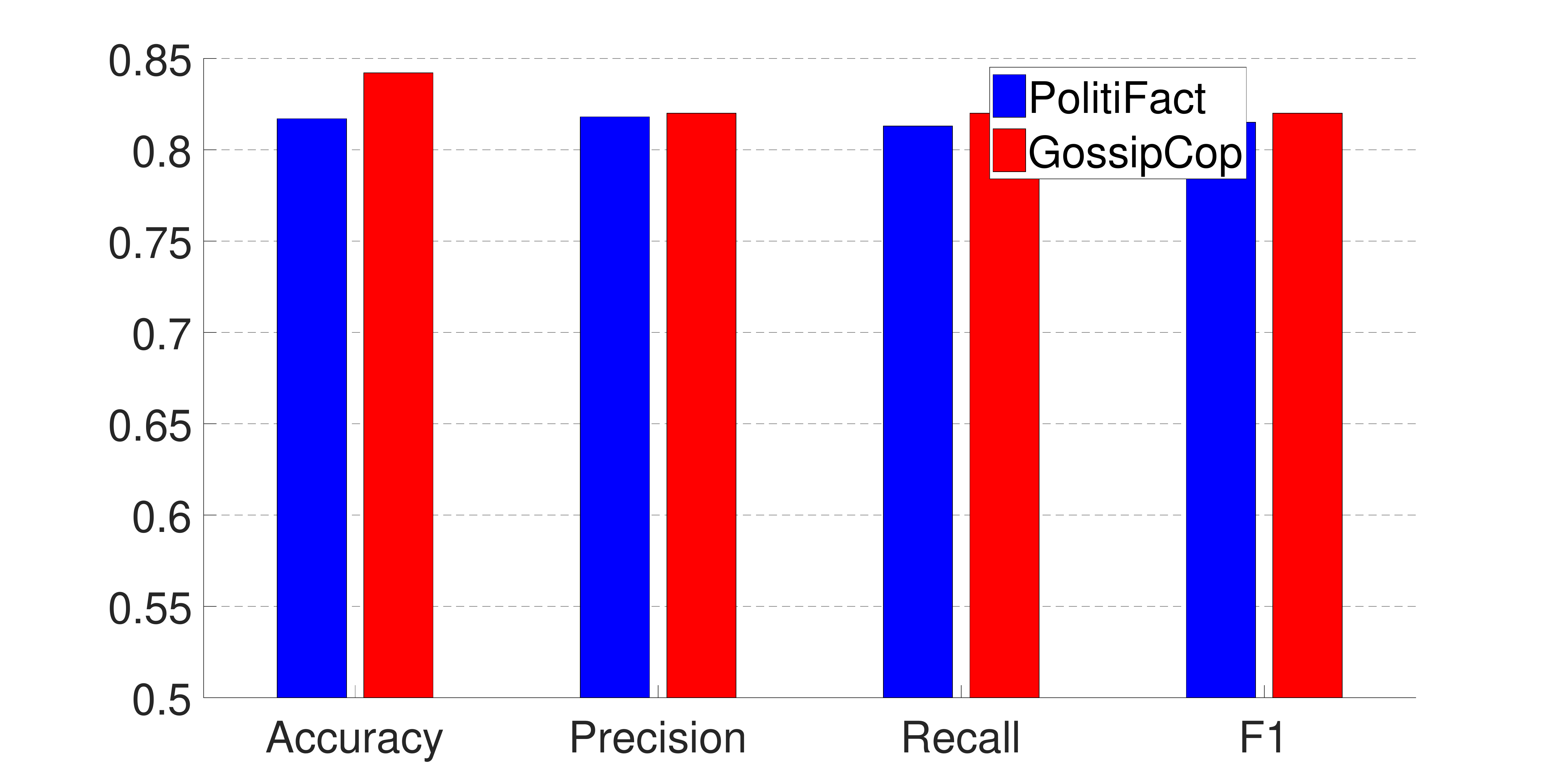}
  \caption{Sample size=300}
  \label{figure_gem_plt_gsp_300}
\end{subfigure}
\caption{Performance of models first trained on the clipped dataset of PolitiFact and then on GossipCop using GEM.}
\label{figure_gem_plt_gsp}
\end{figure*}

\begin{figure*}[ht]
\centering
\begin{subfigure}{.33\textwidth}
  \centering
  \includegraphics[width=\textwidth]{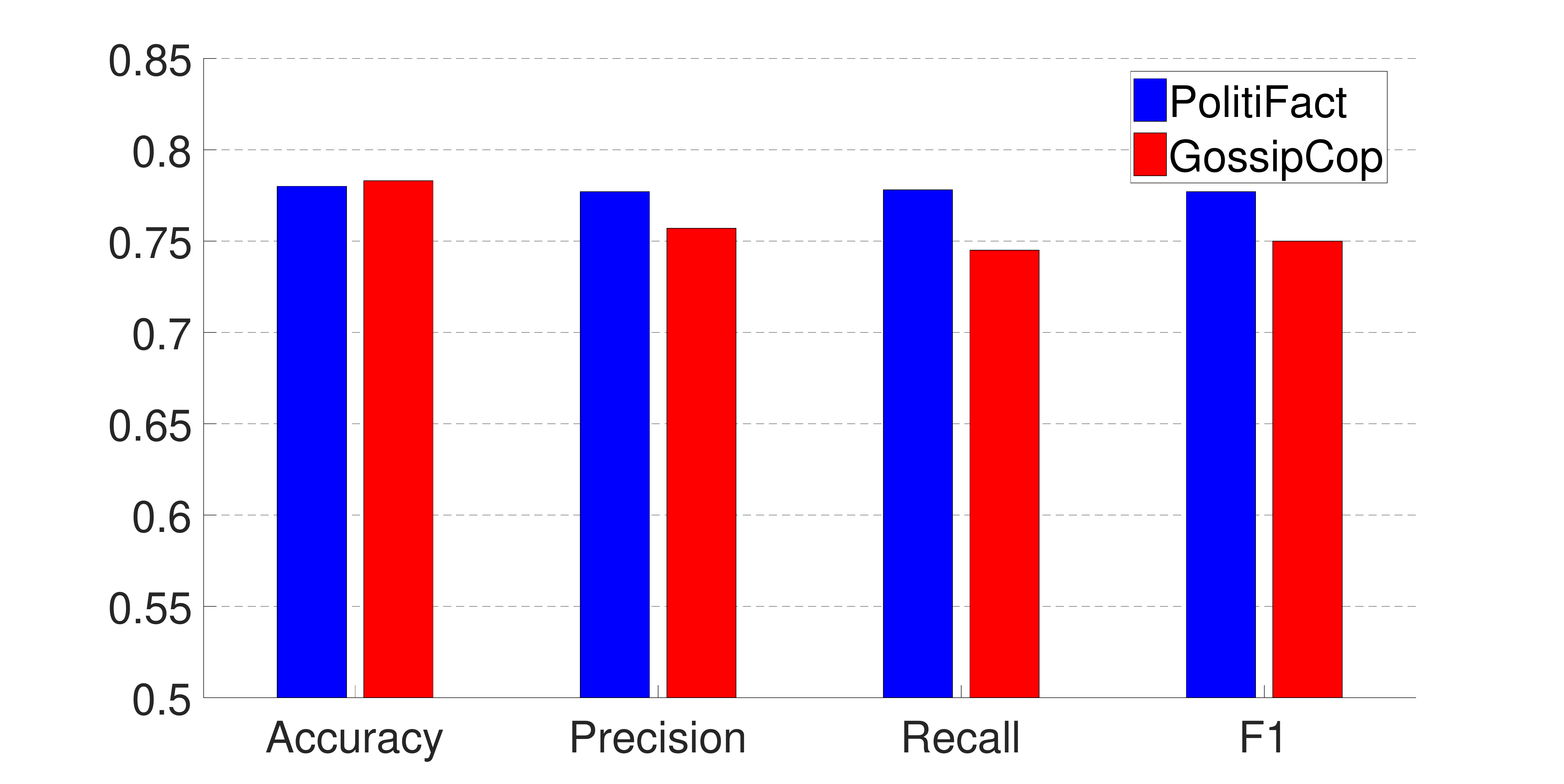}
  \caption{Sample size=100}
  \label{figure_gem_gsp_plt_100}
\end{subfigure}
\begin{subfigure}{.33\textwidth}
  \centering
  \includegraphics[width=\textwidth]{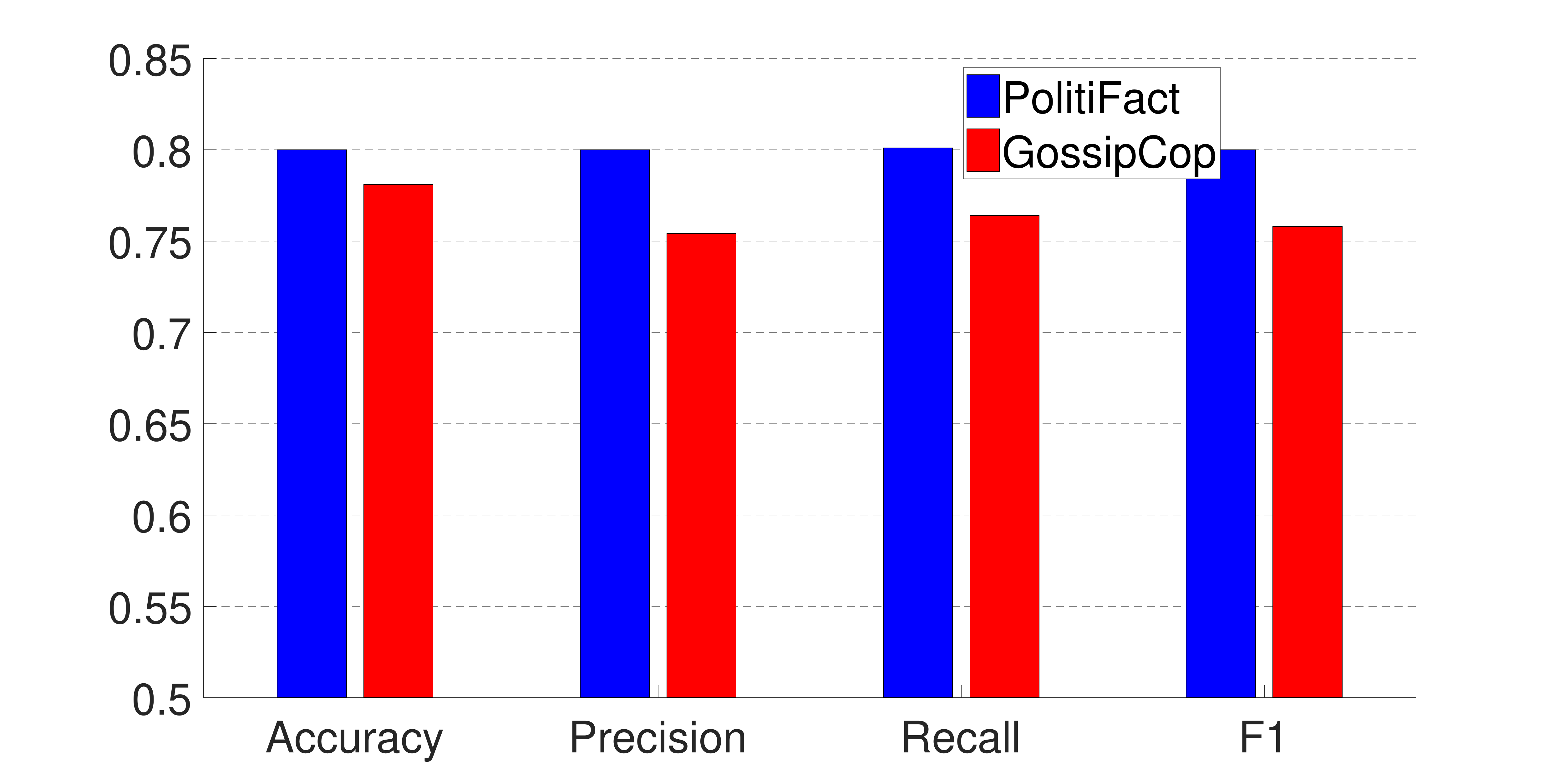}
  \caption{Sample size=200}
  \label{figure_gem_gsp_plt_200}
\end{subfigure}
\begin{subfigure}{.33\textwidth}
  \centering
  \includegraphics[width=\textwidth]{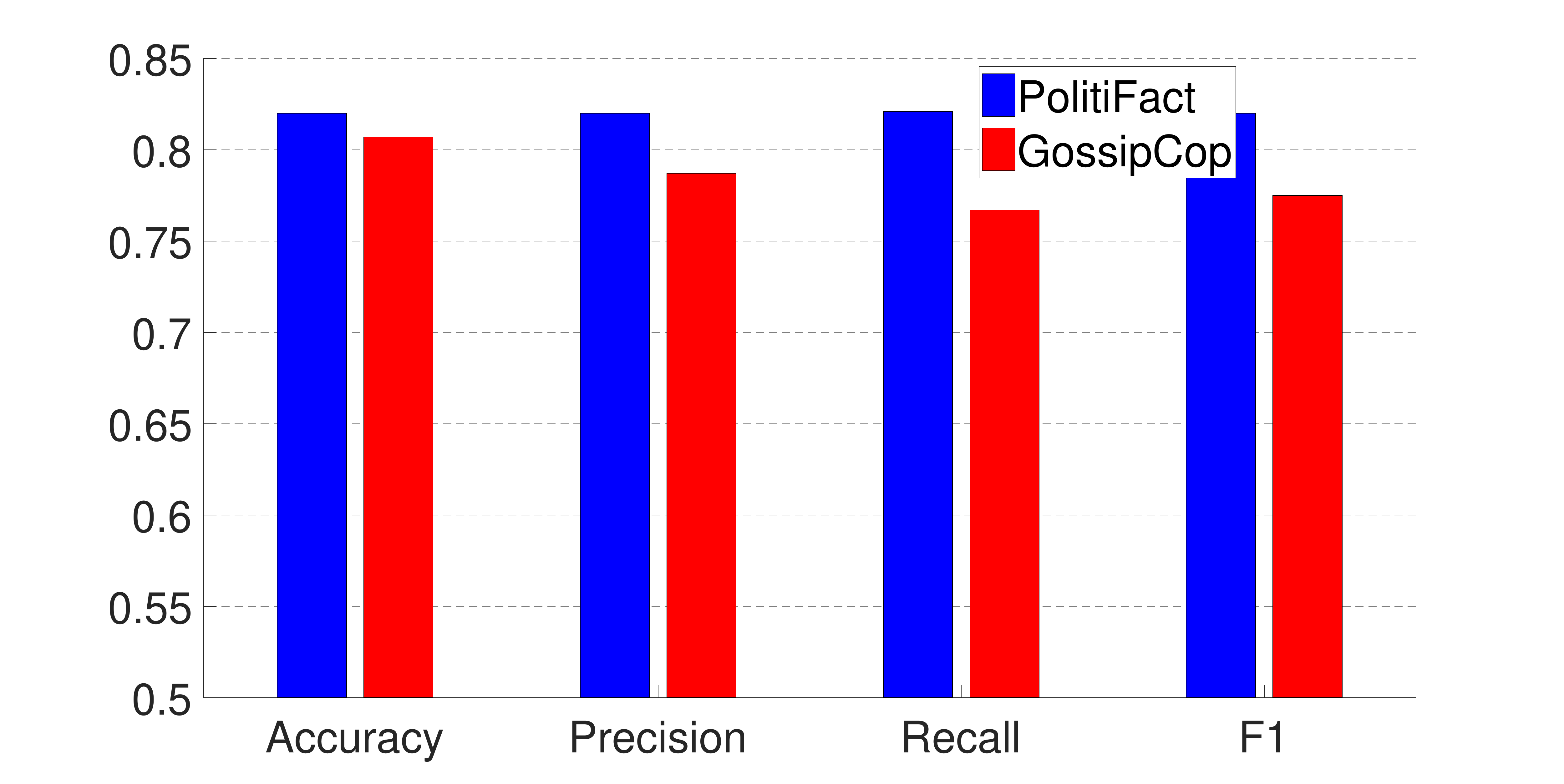}
  \caption{Sample size=300}
  \label{figure_gem_gsp_plt_300}
\end{subfigure}
\caption{Performance of models first trained on the clipped dataset of GossipCop and then on PolitiFact using GEM.}
\label{figure_gem_gsp_plt}
\end{figure*}

In our case, the learning on the two datasets (\(\mathcal{D}_{1}\) and \(\mathcal{D}_{2}\)) are considered as two tasks. When the model learns the first task, it is trained as usual; then during the learning of the second task, we apply GEM and EWC: 

\begin{itemize}
    \item Let \(\theta_{1}\) be the model parameters after the first task, and \(\mathcal{M}\) be the set of instances sampled from the first dataset, then the optimisation problem under GEM becomes:

\begin{align*}
    min_{\theta} \sum_{(G_{i}, y_{i}) \in \mathcal{D}_{2}} loss\left(f(A_{i}^{(k)}, H_{i}^{(k)}; \theta^{(k)}), y_{i}\right)\\
    \text{subject to} \sum_{(G_{j}, y_{j}) \in \mathcal{M}} loss\left(f(A_{j}^{(k)}, H_{j}^{(k)}; \theta^{(k)}), y_{j}\right)\\ 
    \leq \sum_{(G_{j}, y_{j}) \in \mathcal{M}} loss\left(f(A_{j}^{(k)}, H_{j}^{(k)}; \theta_{1}^{(k)}), y_{j}\right)
\end{align*}

    \item Let \(\lambda\) be the regularisation weight, \(F\) be the Fisher information matrix, and \(\theta_{\mathcal{D}_{1}}^{*}\) be the parameters of the Gaussian distribution used by EWC to approximate the posterior of \(p(\theta|\mathcal{D}_{1})\), then the loss function under EWC is:

\begin{align*}
    \sum_{(G_{i}, y_{i}) \in \mathcal{D}_{2}} loss\left(f(A_{i}^{(k)}, H_{i}^{(k)}; \theta^{(k)}), y_{i}\right) + \frac{\lambda}{2}F(\theta-\theta_{\mathcal{D}_{1}}^{*})^{2}
\end{align*}

    Note that when estimating the Fisher information matrix \(F\), we sample a set of instances (\(\mathcal{M})\) and compare the model performance under different sample sizes.
\end{itemize}

In terms of parameters, we test sample size \(|\mathcal{M}|=100, 200, 300\) (all the samples are chosen randomly), and \(\lambda=1, 3, 10, 30, 10^2, 3 \times 10^2, 10^3, 3 \times 10^3, 10^4, 3 \times 10^4, 10^5\) (for EWC only). In addition, since the model architecture has to be consistent during the two phases (\ie first trained on one dataset and then incrementally on the other), the number of pooling layers is set to three.

Figs.~\ref{figure_gem_plt_gsp},~\ref{figure_gem_gsp_plt} and Table~\ref{table_performance_ewc} show the performance of models trained with GEM and EWC (for EWC the results when \(|\mathcal{M}|=100, 200\) are omitted due to the space limit). The results demonstrate that while both methods can achieve a relatively balanced performance over the two datasets, GEM trained models work better than EWC trained models in general. In addition, we have also incrementally trained the model using GEM on the whole dataset, and the performance can be further improved.

Another point worth mentioning is that it requires more fine-tuning during the EWC training process. For example, we need to apply early stopping to ensure balanced results on both datasets when the model is trained with EWC.

\textbf{Efficiency.} In terms of efficiency, the following observations can be made from our experiments on both datasets: (1) compared with the normal training process, training with GEM and EWC requires slightly more time; 
(2) there is no significant difference in training time between GEM and EWC; and (3) the impact of the parameters, \ie sample size and \(\lambda\), on the training time is also not significant.

\begin{table*}[ht!]
    \centering
    \caption{Performance of models trained using EWC (sample size \(|\mathcal{M}| = 300\)).}
    \begin{tabular}{|c|c|c|c|c|c|c|c|c|c|c|c|c|c|c|c|c|}
        \hline
        \multirow{4}{*}{\(\lambda\)} & \multicolumn{8}{c|}{Models first trained on PolitiFact} & \multicolumn{8}{c|}{Models first trained on GossipCop} \\
        & \multicolumn{8}{c|}{and then on GossipCop} & \multicolumn{8}{c|}{and then on PolitiFact} \\
        \cline{2-17}
        & \multicolumn{4}{c|}{PolitiFact} & \multicolumn{4}{c|}{GossipCop} & \multicolumn{4}{c|}{PolitiFact} & \multicolumn{4}{c|}{GossipCop} \\
        \cline{2-17}
        & Acc & Pre & Rec & F1 & Acc & Pre & Rec & F1 & Acc & Pre & Rec & F1 & Acc & Pre & Rec & F1 \\
        \hline
        1 & 0.665 & 0.681 & 0.647 & 0.640 & 0.784 & 0.762 & 0.734 & 0.744 & 0.729 & 0.724 & 0.730 & 0.724 & 0.742 & 0.727 & 0.654 & 0.663 \\
        \hline
        3 & 0.649 & 0.657 & 0.633 & 0.627 & 0.795 & 0.776 & 0.752 & 0.761 & 0.733 & 0.730 & 0.729 & 0.729 & 0.771 & 0.760 & 0.698 & 0.712 \\
        \hline
        10 & 0.677 & 0.684 & 0.662 & 0.660 & 0.777 & 0.758 & 0.724 & 0.735 & 0.731 & 0.732 & 0.728 & 0.728 & 0.766 & 0.746 & 0.701 & 0.713 \\
        \hline
        30 & 0.675 & 0.681 & 0.662 & 0.660 & 0.777 & 0.762 & 0.722 & 0.734 & 0.720 & 0.717 & 0.718 & 0.717 & 0.736 & 0.708 & 0.662 & 0.672 \\
        \hline
        \(10^2\) & 0.683 & 0.687 & 0.671 & 0.670 & 0.780 & 0.764 & 0.720 & 0.733 & 0.720 & 0.718 & 0.719 & 0.717 & 0.740 & 0.719 & 0.657 & 0.667 \\
        \hline
        \(3\times10^2\) & 0.689 & 0.705 & 0.672 & 0.668 & 0.778 & 0.750 & 0.739 & 0.743 & 0.729 & 0.733 & 0.729 & 0.726 & 0.759 & 0.748 & 0.680 & 0.692 \\
        \hline
        \(10^3\) & 0.689 & 0.697 & 0.675 & 0.674 & 0.770 & 0.747 & 0.717 & 0.727 & 0.713 & 0.713 & 0.713 & 0.712 & 0.755 & 0.735 & 0.684 & 0.693 \\
        \hline
        \(3\times10^3\) & 0.695 & 0.703 & 0.681 & 0.680 & 0.770 & 0.745 & 0.711 & 0.722 & 0.718 & 0.720 & 0.718 & 0.717 & 0.733 & 0.704 & 0.660 & 0.669 \\
        \hline
        \(10^4\) & 0.706 & 0.711 & 0.695 & 0.695 & 0.775 & 0.752 & 0.713 & 0.724 & \B 0.718 & \B 0.721 & \B 0.713 & \B 0.712 & \B 0.786 & \B 0.769 & \B 0.732 & \B 0.744 \\
        \hline
        \(3\times10^4\) & \B 0.726 & \B 0.735 & \B 0.714 & \B 0.714 & \B 0.761 & \B 0.739 & \B 0.697 & \B 0.707 & 0.722 & 0.717 & 0.715 & 0.715 & 0.764 & 0.738 & 0.705 & 0.715 \\
        \hline
        \(10^5\) & 0.737 & 0.746 & 0.726 & 0.727 & 0.750 & 0.733 & 0.663 & 0.675 & 0.709 & 0.708 & 0.706 & 0.706 & 0.770 & 0.748 & 0.707 & 0.718 \\
        \hline
    \end{tabular}
    \label{table_performance_ewc}
\end{table*}

\section{Related Work}\label{sec:related}
Detecting fake news on social media has been a popular research problem over recent years. In this section, we briefly review the prior work on this topic. Specifically, similar to~\cite{shu_fake_2017,pierri_false_2019}, we classify existing work into three categories: content-based approaches, context-based approaches and mixed approaches, the first two of which, as suggested by their names, mainly rely on news content and social context to extract features for detection, respectively.

\subsection{Content-based Approaches}
Content-based approaches use news headlines and body content to verify the validity of the news. It can be further classified into two categories: knowledge-based and style-based~\cite{shu_fake_2017,zhou_fake_2018}.

\subsubsection{Knowledge-based Detection}
In order for this type of method to work, a knowledge base or knowledge graph~\cite{nickel_review_2016} has to be built first. Here, knowledge can be represented in the format of a triple: (Subject, Predicate, Object), \ie SPO triple~\cite{noauthor_resource_2017}. Then, to verify an item of news, knowledge extracted from its content is compared with the facts in the knowledge graph~\cite{wu_toward_2014,ciampaglia_computational_2015,shi_fact_2016}. If a triple \((S,\ P,\ O)\) is missing in the knowledge graph, different link prediction algorithms can be used to calculate the probability of an edge labelled \(P\) existing from node \(S\) to node \(O\).
    
\subsubsection{Style-based Detection}
According to forensic psychological studies~\cite{undeutsch_beurteilung_1967}, statements based on real-life experiences differ significantly in both content and quality from those derived from fabrication or fiction. Since the purpose of fake news is to mislead the public, they often exhibit unique writing styles that are rarely seen in real news. Therefore, style-based methods aim to identify these characteristics. For example, Perez-Rosas \etal~\cite{perez-rosas_automatic_2018} train linear SVMs on the following linguistic features to detect fake news: unigrams, bigrams, punctuation, psycholinguistic, readability and syntax features. Other methods that fall into this category include~\cite{horne_this_2017,volkova_separating_2017,wang_liar_2017,potthast_stylometric_2018}.

In addition to textual information, images posted in social media have also been investigated to facilitate the detection of fake news~\cite{jin_novel_2017,yang_ti-cnn_2018,wang_eann_2018,zhou_safe_2020}.

\subsection{Context-based Approaches}
Social context here refers to the interactions between users, including tweet, retweet, reply, mention and follow. These engagements provide valuable information for identifying fake news spread on social media. For example, Jin \etal~\cite{jin_news_2016} build a stance network where the weight of an edge represents how much each pair of posts support or deny each other. Then fake news detection is based on estimating the credibility of all the posts related to the news item, which can be formalised as a graph optimisation problem. 

Tacchini \etal~\cite{tacchini_like_2017} propose to detect fake news based on user interactions, \ie users who liked them on Facebook. Their experiments show that both the logistic regression based and the harmonic Boolean label crowdsourcing based methods can achieve high accuracy.

Unlike the above supervised methods, an unsupervised approach is proposed in Yang \etal~\cite{yang_unsupervised_2019}. It builds a Bayesian probability graphical model to capture the generative process among the validity of news, user opinions and user credibility.

Note that propagation-based approaches as mentioned in the introduction also belong to this category.

\subsection{Mixed Approaches}
Mixed approaches use both news content and associated user interactions over social media to differentiate between fake news and real news.

Ruchansky \etal~\cite{ruchansky_csi_2017} design a three-module architecture that combines the text of a news article, the received user response and the source of the news: (1) the first module takes the user response, news content and user feature as the input, and trains a Recurrent Neural Network (RNN) to capture temporal representations of articles; (2) the second module is fed with user features to generate a score and a low-dimensional representation for each user; (3) the third module takes the output of the first two modules and trains a neural network to label the news item.

Zhang \etal~\cite{zhang_fakedetector_2018} propose to use a pre-extracted word set to construct explicit features from the news content, user profile and news subject description, and meanwhile use a RNN to learn latent features, such as news article content information inconsistency and profile latent patterns. Once the features are obtained, a deep diffusive network is built to learn the representations of news articles, creators and subjects.

Shu \etal~\cite{shu_beyond_2019} use the tri-relationship among publishers, news articles and users to detect false news. Specifically, non-negative matrix factorization is used to learn the latent representations for news content and users, and the problem is formalised as an optimisation over the linear combination of each relation. Multiple machine learning algorithms are tested to solve the optimisation problem, and the results demonstrate its effectiveness.

In addition to the above work, a few recent papers have started to work on explainability, \ie why their model labels certain news items as fake~\cite{popat_declare_2018,shu_defend_2019,lu_gcan_2020}.

\section{Conclusions and Future Work}\label{sec:conc}
The prevalence of fake news over social media has become a serious social problem. In this paper, we propose a propagation-based method approach for fake news detection, which uses GNNs to distinguish between the different propagation patterns of fake news and real news over social networks. Even though the method only requires a limited number of features obtained from the social context, and does not rely on any text information, it can achieve comparable or superior performance to state-of-the-art methods that require syntactic and semantic analyses.

In addition, we identify the problem that GNNs trained on a given dataset may not perform well on new data where the graph structure is vastly different, and direct incremental training cannot solve the issue. Since this is similar to the catastrophic forgetting problem in continual learning, we propose a technique that applies two popular approaches, GEM and EWC, during the incremental training, so that balanced performance can be achieved on both existing and new data. This avoids re-training on the entire data, as it becomes prohibitively expensive as data size grows.

For future work, we will investigate whether, to some extent, the catastrophic forgetting phenomenon in this case can be mitigated by the choices of features---include more features, or find ``universal" features that work well despite the different graph structures.

\bibliographystyle{ACM-Reference-Format}
\bibliography{references}

\end{document}